# Body Dust: Miniaturized Highly-integrated Low Power Sensing for Remotely Powered Drinkable CMOS Bioelectronics

Sandro Carrara[#] and Pantelis Georgiou*

[#] Integrated System Laboratory, EPFL, Lausanne, Switzerland and *Centre for Bio-Inspired Technology, Imperial College London, UK

Correspondence to: sandro.carrara@epfl.ch

The aim of this paper is to introduce current advances in technology that could enable the development of fully drinkable and autonomous bio-electronic CMOS sensors in the form of dust particles, capable of identifying the source of a disease by targeting a specific region in organs and tissue such as a tumor mass and automatically sending diagnostic information wirelessly outside the body. We call this swarm of sensing dust particles 'Body Dust'. A diagnostic system in the form of Body Dust would need to be small enough to support free circulation in human tissues, which requires a total size of less than 10 μm$^3$, in order to mimic the typical sizes of a blood cell (e.g., red cells have the diameter around 7 μm). Whilst with present state-of-the-art in CMOS technology, this requirement in terms of size is currently un-feasible, recent research has advanced technology such that we can begin to work towards such an approach. Therefore, we present here the current limits of CMOS technology as well as the challenges related to the development of such a system. Towards this goal, this article presents the theoretical feasibility to obtain the first ever-conceived sub-10 μm Bio/CMOS integrated circuit with biosensing capability to provide diagnostic telemetry once self-located in human tissue.

CMOS design | Heterogeneous Systems | Biosensors | Analog Design Telemetry | Cancer Diagnostics | Carbon Nanotubes | Nano-Platinum

In the last 10 years, the field of microelectronics has progressed significantly due to the introduction of nano-scale CMOS technology for integrated circuits (IC), guaranteeing robustness and reliability though novel circuit-design techniques (*1*). This amazing progression has also been seen in the design of CMOS interfaces for biosensing, allowing realisation fully integrated micro-scale sensing systems. (*2*). Alternatively, the development of nano-fabricated devices has paved the way of new kinds of biosensing (*3*), while offering the possibility of fusion with also sample preparation and/or amplification to create fully autonomous sensing systems (*4*). By pushing forward this trends in CMOS and nano-fabrication of sensors, we can begin to conceptualise the development of an extremely integrated system the same size as a human cell and which also contains all the electronics for in-body sensing and wireless telemetry of data and whose form factor is so small that it could exist as a swarm of dust particles in a diagnostic solution that could be drunk by a patient requiring diagnosis.

Working towards this goal, in this paper we introduce the theoretical feasibility of the first ever conceived sub-10 μm Bio/CMOS integrated circuit with biosensing capability useful to provide diagnostic telemetry once self-located in a human tissue after being drunk by the patient. The core idea is here to propose the design of a fully packaged CMOS cube as illustrated in Figure 1, which integrates the most compact, in terms of area, integrated circuits ever built. This CMOS cube has to be capable of providing diagnostics through sensing of specific molecules, data-telemetry, low-power consumption, and biocompatibility. An adapted energy-harvesting system to provide enough power is also required. We also envisage the use of the most-robust, smallest, and commercially available CMOS nodes (e.g., 5 nm) as well as the addition of special functions to the CMOS architecture for dealing with post-processing heterogeneous integration of further nanostructures.

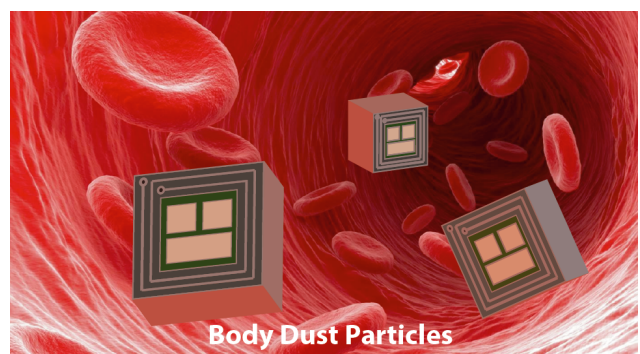

**Fig. 1**: Illlustration of the conept of CMOS Body Dust. Shown are diagnostic sensing cubes of 10 μm$^3$ comparable with the size of red blood cells.

These nanostructures are typically required to form sensors with increased sensitivity for diagnostics (*5*). Once realized, this bioelectronics system will have the form of Body Dust due to its μm-scale size and be capable of real-time diagnostics inside human tissues as well as in organs. Towards realising such a system, this research paper provides a comprehensive investigation of the options along the design path of such a drinkable bio-electronic dust system that is capable of self-identifying a target body-region source of a disease (e.g., a tumor mass) and automatically sending wirelessly diagnostic data outside the body.

## System Concept

A diagnostic system with the mentioned features is envisaged to be of the form a myriad of small particles, each one with a total size of less than 10 μm$^3$. This is required for passing the gut-wall barrier by the mechanism of cells internalization (as well as blood cells) and, therefore, diffusing in the human tissues. Once diffused in the body, these free circulating diagnostic particles need to have the capability of self-targeting a specific body region (e.g., a tumor mass). Once onto the body-mass, these particles need to start providing diagnostics through telemetry. The telemetric diagnostics needs to be at level of specific molecules in order to monitor the trend of the disease (e.g., by measuring specific biomarkers or metabolites) as well as to provide feedback for personalized medicine (e.g., by measuring specific therapeutic compounds). Toward that aim, the research path needs to develop along the following main branches: (i) demonstrating the feasibility of designing the most compact CMOS IC allowing in any case all the required functions to support telemetric diagnostics as well as any function required for its initial setting; (ii) demonstrating the feasibility of designing of a proper system to address the powering issues arising at such small scale of integration, (iii) showing monolithic integration with the right biomaterials in order to assure specificity for the diagnostics aims, (iv) showing monolithic integration with the right nanomaterials in order to assure enough sensitivity for the diagnostics aims; (v) demonstrating the design of a biocompatible packaging to avoid any inflammatory reaction by the human body; (vi) assuring a further coating-function to target the specific body mass; (vii) assuring another coating-function to allow the internalization in M cells and dendritic cells, special cells devoted to internalise big bio-structures (e.g., liposomes with sizes up to 2.5 μm), and, thus, to allow passing through the barrier of the gut-wall. With the present state-of-the-art in CMOS technology, targeting a total size of less than 10 μm$^3$ looks to be un-feasible. However, recent findings obtained over the last very few years open the possibility for realizing such a CMOS cube we propose in form of Body Dust. In fact, compact CMOS circuits for glucose sensing at a size of 0.36-mm$^2$ have been realised in 2012 by using the 0.13-μm CMOS process (6) and similar designs could be further reduced in size by both further simplifying the design and then fabricating in smaller nodes (e.g., with CMOS 5-nm or smaller). Micro-coils with lateral sizes from 200 μm (7) down to 50 μm (8) have been published in 2016 for successfully powering in implantable micro-stimulators. Biosensors realized on CMOS dies with radii of the working electrodes down to 2.5 μm have been published in 2014 and successfully tested for glucose detection thank to platinum nanostructures (9). We successfully realized, tested, and published in 2016 a biocompatible packaging obtained with an epoxy-resin to protect telemetric diagnostic devices fully implanted in mice (10). The use of affinity ligands for specific uptake to targeted disease-cells is one of the proposed strategy to assure active targeting in anticancer therapies (11). The mechanism of internalization provided by M cells and dendritic cells is a well-known phenomenon (12) that allows active transport through cells-layers of bio-structures with also size in few μm (e.g., liposomes present size ranging from 0.025 μm to 2.5 μm) (13).

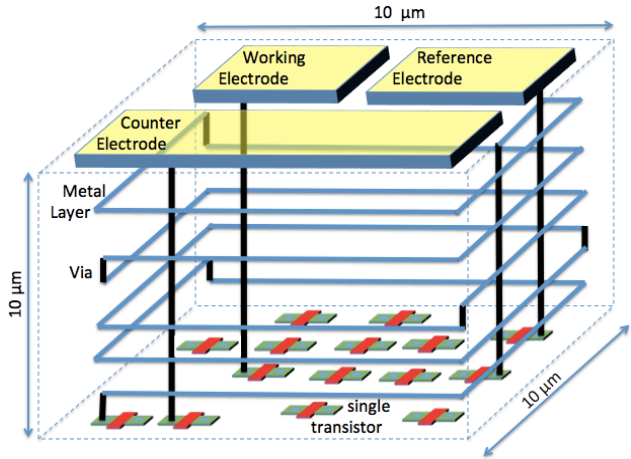

**Fig. 2**: Concept of the CMOS cube for Body Dust with an electrochemical sensor on top of the last metal-layer and powering antennas realised by using all metal-layers.

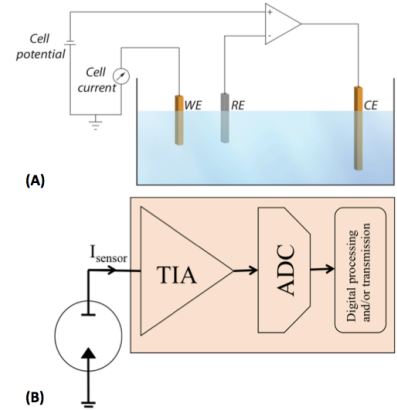

**Fig. 3**: (A) Diagram of a three-electrode electrochemical cell with a potentiostat; and (B) the conceptual diagram of a readout IC to read the sensor current (reprinted from (25)).

Moreover, it is also well-known that the gut permeability is usually augmented in Crohn's disease-patients and ulcerative colitis (14).

Considering the research described shows that mechanisms are possible for realisation of Body Dust particles, we therefore propose in this paper the concept of a CMOS cube of 10 μm$^3$ as shown in figure 2 as first prototype in order to investigate all the difficulties related to developing a system with such an extremely-small size for an active biosensing device. For such a demonstration, we will start, first, by introducing the state-of-the-art about the main issues related to this new design: integrated potentiostats, powering systems, nano-biosensors on chip and packaging. Then, we will propose an original design for the proposed system, including heterogeneous integration of sensors and packaging. The paper closes with discussion on the proposed design and conclusions.

### State-of-the-art: the potentiostats

Amperometric methods are one of the most widely adopted and commonly used approaches for biosensing applications. Their mode of operation relies on the detection of signals in the form of an electric current that is proportional to the concentration of the target biomolecule. Initiating a redox reaction normally generates the current.

As shown in Figure 3 (A), to initiate the reaction, a cell potential (or voltage) is normally applied to a reference electrode (RE) to generate a cell current measured through a working electrode (WE). A circuit called the potentiostat, is typically required to apply this voltage and measure the current. Fully integrated potentiostats aim to combine surface electrodes and instrumentation working as a lab-on-chip device to detect various electrochemical species. As shown in Figure 3 (B), the instrumentation to read the generated current typically requires a transimpedance amplifier (TIA) to convert the cell current to a voltage, which is then sampled by an analogue to digital converter (ADC) for data transmission. In implantable applications, in addition to having low-power, important parameters of the potentiostat are the input resolution and dynamic range, which defines the range of current that can be sensed as well as the smallest detectable current. An ADC is sometimes integrated with the potentiostat IC in order to provide a digital output that can be directly transmitted wirelessly.

**Table 1**: Performance of some fully integrated potentiostats

| Method | Input Resolution- Max input (Dynamic range, dB) | Power (μW) | Digital output | Reference |
|---|---|---|---|---|
| Resistive TIA | 22 pA- 1 μA (93) | 90 | No | (15) |
|  | 330 pA- 40 μA (101) | 495 | No | (16) |
| Sigma-delta | 100fA- 0.5 μA (140) | 3.4 |  | (17) |
|  | 56 pA- 750 nA (82) | 72.5 | Yes | (18) |
|  | 0-32 μA (60) | 25 | Yes | (19) |
|  | 1pA- 1μA (120) | 42 | Yes | (20) |
| Switch-Capacitor OpAmp integrator | 60pA- 10μA (124) | 3000 | No | (21) |
|  | 0.47pA-20μA (156dB) | 9300 | Yes | (22) |
|  | 240 pA- 110 nA (53) | >10000 | Yes | (23) |
| Current-mode | 8.6 pA- 350 nA (92) | 4 | No | (24) |

A summary of the state-of-the art of potentiostats found in literature with their readout circuits and achieved specifications is presented in Table 1. The circuit configurations for readout usually fall into one of the following four categories: (i) resistive feedback transimpedance amplifier (TIA), (ii) sigma-delta modulator, (iii) switched-capacitor op-amp integrator, and (iv) current-mode readout. The resistive feedback TIA is the simplest circuit to implement as it measures in continuous time and it provides good input current resolution. The circuits in the sigma-delta modulator class have the widest input dynamic range without the need for changing the gain. The switched capacitor-integrator converts the biosensor current response into voltage, which then goes through a programmable-gain-amplifier to support high dynamic range. Finally the current-mode circuits use current mirrors and conveyors to produce a copy of the sensor current and to amplify it before quantising it with the ADC.

In terms of achieving minimum area through integrated circuit design, a CMOS instrumentation system-on-chip for glucose sensing has been successfully realised in 2012 with a lateral size around 600 µm (total area of 0.36-mm²) (*6*). This IC has been fabricated by using a 0.13-µm CMOS process, while further simplified architectures fabricated by using smaller CMOS nodes (e.g., with CMOS 5-nm) could reduce the area further. Another example is a prototype of a time-to-digital converter fabricated in 65nm CMOS, achieving a lateral size of only 245 µm (total area of 0.06-mm²) (*26*), while a brain stimulator has been very recently proposed with lateral sizes of 220 µm x 180 µm (total area less than 0.04 mm²) and fabricated in 0.18 µm CMOS technology (*7*).

### State-of-the-art: micro-fabricated coils

A coil with sizes of 200 µm x 200 µm was very-recently fabricated off-chip and then used to power one of the previously mentioned CMOS IC designed with very small lateral sizes. The power delivered to the system was 0.016 mW (*7*), which is close to the power required by quite complex CMOS potentiostats for multiplexed detection of human metabolism (*27*). On the other hand, coils with smaller size, down to 50 µm, have been proposed in 2016 for powering implantable brain stimulators (*8*). In this second approach, the micro-coil was fabricated in two versions: a first version was obtained by a copper trace (10 µm wide × 2 µm thick) directly fabricated on a silicon substrate and passivated with 300 nm of $SiO_2$. A second version was indeed realized by carefully bending an ultra-fine wire of copper with only 50-µm in diameter. The realized coils have a profile of 50 µm × 100 µm and a length of 2000 µm. The coil realized with the micro-wire supported a stronger current even though was tough to get very sharp structures as obtained with the coil micro-fabricated directly on silicon. The power delivered by these new micro-coils compare favourably to power levels around 0.5 mW, thus well enough for also powering quite complex CMOS full-systems for multiplexed detection of human metabolism (*27*).

### State-of-the-art: nanosensors and biosensors

To provide diagnostics on specific molecules (e.g., cancer biomarkers, human metabolites, anti-cancer therapeutic compounds), we can transfer specific recognition functions to the working electrodes shown in Figure 2. This is typically obtained by incorporating probe-proteins that specifically recognise the target-molecules with respect to the sensing aim. For example, Antibodies (*28*) or Aptamers (*29*) are extensively used for detecting cancer biomarkers, oxidases (a class of enzymes) are proposed for several human metabolites (*30*), cytochromes P450 (another class of enzymes) are suggested for anticancer therapeutic compounds (*31*).

In many cases, protein based amperometric biosensors face as limiting factor the electron transfer rate toward the working electrode. To improve electron transfer rate, many studies propose the characterization of protein sensing-systems working on nanostructured electrodes. For example, multi-walled carbon nanotubes (MWCNT) are used for structuring carbon-paste based (*32*) as well as gold electrodes (*33*). The typical obtained surfaces are similar to that shown in the following Figure 4, where highly dense bundles of carbon nanotubes were formed on the sensing surface by drop casting solutions with mono-disperse nanotubes.

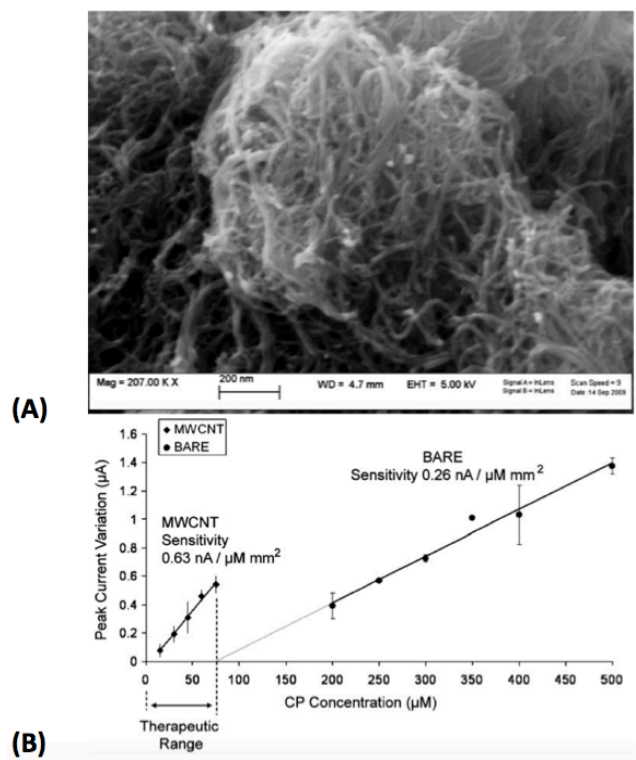

**Fig. 4**: (A) Multi-walled carbon nanotubes on sensing electrodes; and (B) detection of a chemotherapy compound with or without multi-walled carbon nanotubes (reprinted by (*31*))

Such surfaces with highly dense bundles of carbon nanotubes turned out to be extremely powerful in terms of sensor performance for both enzyme-mediated and direct detection of human metabolites. For example, in case of anti-cancer drugs, it was possible to achieve sensitivity in the right pharmacological range of concentrations only thank to the integration of such nano-materials, as shown in Figure 4 (B) for the case of a very well-known cancer chemotherapy compound (the cyclophosphamide, CP). Of course, different nanomaterials transferred or growth on metal surfaces return different sensing performance. The glucose detection has been differently improved by different nanomaterials as per data in Table 2. Several approaches have been proposed along the last years to incorporate Carbon Nanotubes (CNT) onto the sensing electrodes. CNT may be integrated on metal surfaces by direct CVD (Chemical Vapour Deposition) growth (*42*), micro-spotting (*43*), or electrochemical deposition (*41*). Graphene like nanostructures are integrable on metals again by direct CVD growth too (*44*), while nano-platinum is usually obtained by electrodeposition (*38*). Thank to platinum nanomaterials easily integrated on electrochemical CMOS devices, biosensors directly realized on top of the CMOS dies with radii of the working electrodes down to 2.5 µm have been published in 2015 and successfully tested for glucose detection (*9*). With this kind of nano-structuration, electrodes with sizes of only 5 µm in diameter were found with a sensitivity increased by a factor larger than 400 times (*9*).

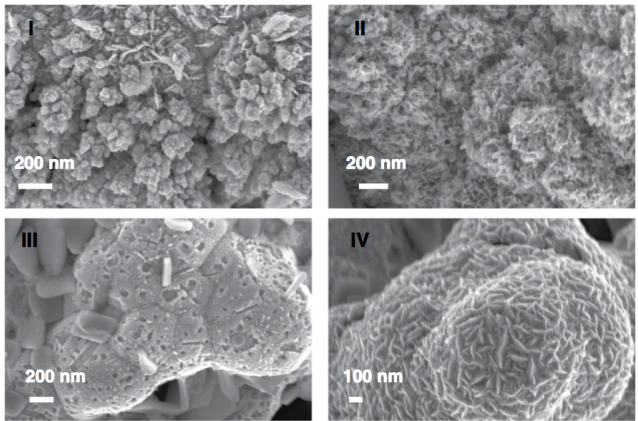

**Fig 5:** Morphology of electrodeposited Pt nanostructured films (reprinted from (*38*).

With the electrochemical deposition, we have successfully obtained sensors with high-performance by also transferring nano-Pt structures (Figure 5), by reaching a maximum sensitivity of 51.6 µA/(mM cm2) (*38*). With this nano-Pt, we have detected several metabolites of extreme interest in biomedical applications, including but not limited to glucose (as shown in Figure 6) and potassium (*5*) in undiluted cell culture media. Over the recent years, we have detected several other endogenous and exogenous molecules related to human metabolism and cancer diagnostics by combining nanostructured sensors and probe-proteins. For example, we have successfully obtained the best ever realised aptamers based biosensors for the Prostate (cancer) Specific Antigen, the PSA (*3*) as well as an antibodies based biosensors for another important cancer marker, the grow factor VEGF (*45*). We have detected several anticancer chemotherapy compounds, including cyclophosphamide, ifosfamide, ftorafur, and etoposide, both in buffer samples and in human serum, within the pharmacological ranges (*46*).

### State-of-the-art: nanosensors on chip

In some cases, the designed IC needs to directly integrate nanostructured biosensors in order to provide high performance on real human samples. In such cases, a possible design-solution is the opening of the IC die to realise the sensing electrodes by metal evaporation directly on top of the IC, as shown by the SEM image in Figure 7. In this manner, the sensing electrodes are directly exposed to the bio-interface as well as in direct electrical contact to the underneath CMOS metal layers. This solution has been proposed to enables a close positioning of the electronics to the sensing chambers. Figure 8 shows a biochip designed for DNA detection, where the location of the CMOS IC is extremely close to the sensing interface, assuring the best possible signal-to-noise ratio.

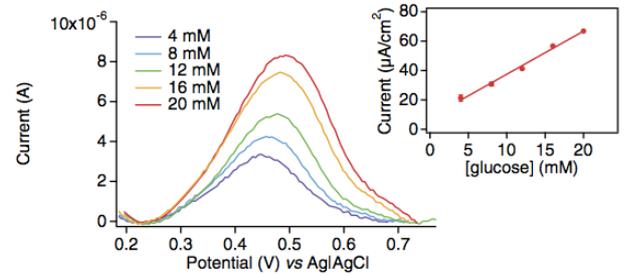

**Fig. 6:** Amperometric detection by cyclic voltammetry of glucose in cell media (reprinted from (*38*)).

This kind of full integration of biosensing devices and CMOS architectures not only improves signal/noise ratio, and accordingly provide smaller limit-of-detections (LOD), but also address the need of smaller biochips for portable, wearable, implantable, or injectable electronics for human health continuous monitoring (*27*). Over the last years, we have successfully achieved several milestones toward a deep integration of heterogeneous structures for sensing application with CMOS designed potentiostats. For example, our group achieved for the first time the fabrication of carbon nanostructures with sufficiently good yield at 450ºC (*44*), a relatively lower temperature compatible for integration with CMOS technology (*48*), since IC dies and their adhesive materials are usually start melting at temperatures as high as 500 °C (*49*). We have also demonstrated an alternative method to transfer nanomaterials by micro spotting. However, it turned out that this method is limiting the sensing performance (sensitivities about 0.46 µA/mM cm2) due to the presence of required polymers (*43*). While chemical vapour (*44*) or electrochemical (*41*) depositions return the best sensitivities in transferring carbon nano-materials. The biosensing performance of these systems have been improved by also integrating MWCNT by drop casting (*41*) or electrodeposition (*30*). By using these approaches, we have succeeded in monitoring several human molecules by integrating proper proteins, e.g. enzymes from the families of oxidases or cytochromes P450. We have detected endogenous metabolites, e.g, glucose and lactate (*27*), as well as exogenous therapeutic compounds, such as etoposide and mitoxantrone (*41*), or etodolac (*50*). Fully implantation in mouse has been successfully verified too with the detection of glucose and paracetamol (*51*), as models for the detection of both endogenous biomarkers and exogenous therapeutic drugs. We have also significantly contributed to the integration of CMOS ICs with these kinds of protein-based electrochemical biosensors, especially for implantable devices. Figures 9 shows two fully integrated multi-panel systems we have designed and realized very few years ago. Both the systems present several biosensors and also include a pH sensor and a temperature sensor.

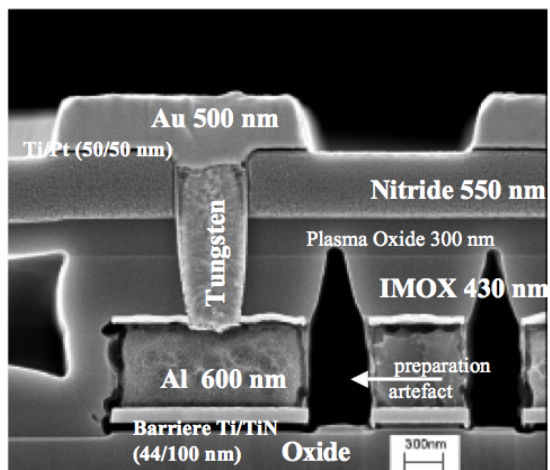

**Fig. 7:** Sensing electrodes directly created on top of the last metal layer (from (*47*))

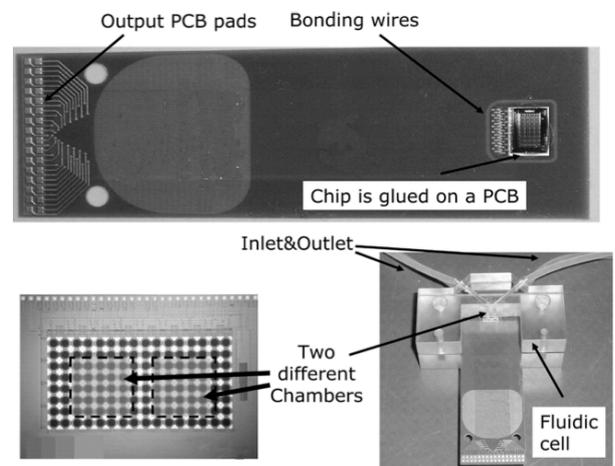

**Fig.8:** Integrated BioChip with 128 couples of electrodes for DNA detection (from (*52*))

### State-of-the-art: biocompatible packaging

In case of implantable or injectable electronics, issues on biocompatibility take a major role, as the hosting body needs to be protected by any inflammatory reaction. In this respect, several tens of years of research and development about implantable devices then successfully introduced into the market (e.g., pacemakers or systems for deep brain stimulation) have proposed several solutions to obtain biocompatible packaging. For example, fully sealed packaging has been proposed in biocompatible silicone (*53*), while semi-permeable membranes has been proposed for sensing aims by using polycarbonate (*54*) or epoxy-resin (*55*). The last two are easy methods to cover a diagnostics system since they allow human metabolites to penetrate the packaging and reach the sensing electrodes of the system. Figure 9 shows these two methods of building biocompatible packagings that we have used in past to realise implantable biosensing chips with biocompatible covers. The biocompatibility of the membranes obtained with such methods has been already tested and published several times in literature, as well as we have confirmed with experiment on mice (*10*).

### System Design

As we have seen in all the previous sections about the present state-of-the-art in CMOS technology for heterogeneous integration toward fully integrated biosensing systems, many already achieved developments are now pointing toward the direction of extremely small-scale devices and in the next 5 or 10 years the technology will be ready to bring to the market Body Dust under form of CMOS active sensing cubes with later sizes of about 10 μm³. In fact, even though it seems that fully integrated biochips are now still limited to lateral sizes of about 200-500 μm, we have already seen in literature as well as partially demonstrated with our previous works already published all the knowledge and prior art we need to push now the present limits well-below 100 μm and, therefore, investigate all the design-challenges required for a further step toward CMOS diagnostics systems with lateral sized below 10 μm. This and the following sections are then going to introduce how to design such a Body Dust with self-forming, self-localization, and diagnostics capability.

### System Design: the sensor frontend

As the first step in design, and specifically for the system-on-chip which will perform the sensing, the following blocks are considered: A **3-electrodes set** containing counter, working and reference electrodes; a transimpedance amplifier, a potentiostat, a reference block, and a current controlled ring oscillator.

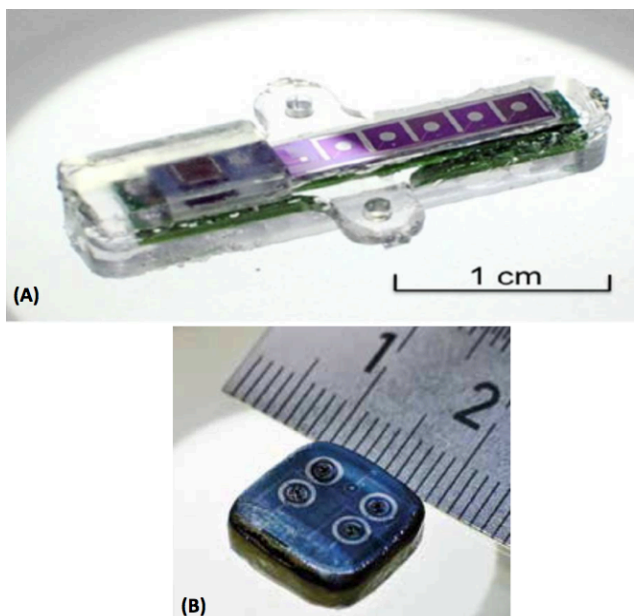

**Fig. 9:** ICs fabricated with multi-target sensors within biocompatible packages: (A) reprinted from (*27*), and (B) reprinted from (*41*).

A possible complete system architecture for this system on chip is shown in Figure 10. Specifically, it is made up of the following:

- A **transimpedance amplifier (TIA)** connected to the working electrode to convert the sensor current flowing through the working electrode into a voltage. This circuit needs to be high gain, low noise (sub pA/√Hz) and have a high dynamic range (>120dB) capable of detecting bi-directional sensor currents from 1pA-100 $\mu$ A. The bandwidth should allow operation of up to 100KHz to carry out cyclic voltammetry and chronoamperometry. Several topologies including those shown in Table 1 are considered in this design to achieve the best-required specifications.
- A **potentiostat** is designed to be connected to the reference electrode and counter electrode. This circuit needs to provide a stable bias to the reference electrode and have sufficient bandwidth to allow cyclic voltammetry. An opamp-based approach is used with high gain (>90dB) and bandwidth (>100KHz), and is designed rail-to-rail so that it can supply voltages to achieve the maximum dynamic range.
- A **Modulator** implemented as a current **controlled ring oscillator** to convert the sensed current from the TIA to a frequency-modulated signal, which drive the transmission coils through backscattering.

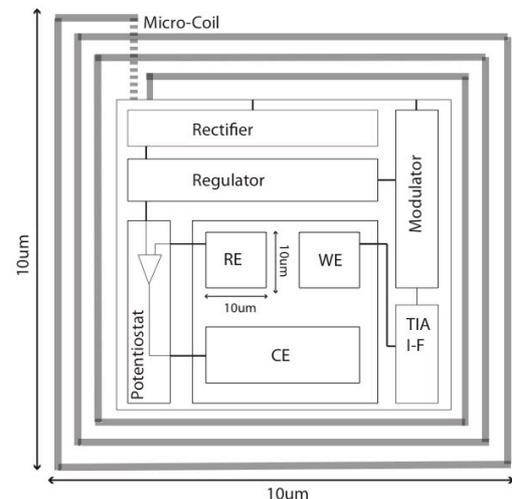

**Fig. 10**: System architecture of self-forming CMOS SoC.

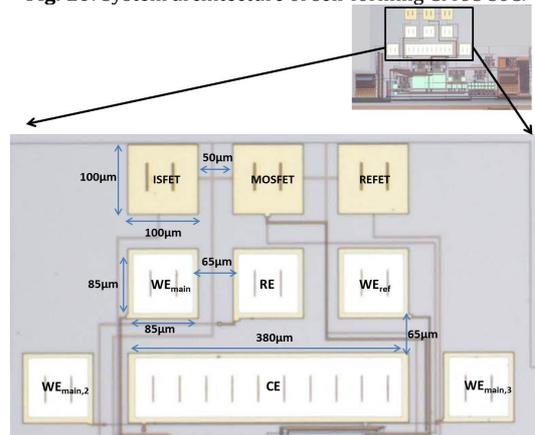

**Fig. 11:** Aluminium electrodes created on the surface of a 0.35μm CMOS chip, from (*22, 56*)

- An on-chip receiver and **power rectifier** to extract power and generate a **regulated** supply form the signal received by the coil. This should be designed to generate a supply between 0.5-1.2V.

We have already designed a similar CMOS architecture and related simulations have demonstrated the possibility to realize such a design within a chip area of about 111,64 μm² (*57*), which fits well with the aim of this paper. On top of such a CMOS die, an array of aluminium electrodes is formed on the top CMOS surface to realise the working, reference and counter electrodes, which allows the platinum deposition. These acts as the sensor interface for the diagnostic ICs. As shown in Figure 11 and referred to one of our previous works (*56*), aluminium is the standard metal typically used in a CMOS process and the electrodes are obtained using the top metal of the CMOS process.

CMOS microchips are typically passivated on the top surface with an insulating layer of silicon dioxide to protect from humidity and damage. This layer is etched at fabrication to allow exposure of top metal aluminium for the purpose of forming bond pads to the chip usually enabling wire bonding of gold wires in order to interface to the integrated circuits on the chip by using external electronics. We use the same pad layer opening to form our top electrodes. This is achieved by designing the electrode-area by using top metal, and then defining this area as a pad layer. During fabrication, all areas defined as pad layer is etched until top-metal which is used as an etch stop. The result is the formation of aluminium electrodes of similarly in the structure shown in Figure 11. These aluminium electrodes are then directly connected to the integrated circuits by the underneath metal layers to form a complete monolithic solution.

*Integration of remote powering coil*
For the integration of the remote powering micro-coil, two different strategies are considered. First, a coil micro-fabricated is designed and realized independently by the CMOS circuits by depositing metals on silicon by using conventional lithography. For this first approach, we propose a fabrication process we have already tested. In fact, we have already successfully realized, tested, and published in 2014 a coil micro-fabricated directly in silicon and realized in the clean room of EPFL (*58*), as shown in Figure 12. This coil was realized with a single-layer deposition by using a mold with thickness from 60 μm and it is capable to receive up to 8.7 mW. In particular, thin layers of chromium (20 nm) and gold (100 nm) are evaporated by Joule effect on the wafer. Then, a layer of 60 μm Ordyl dry film is laminated on the wafers as negative photoresist. Trenches are opened in the Ordyl film and copper electro-plating is performed to deposit the inductor traces.

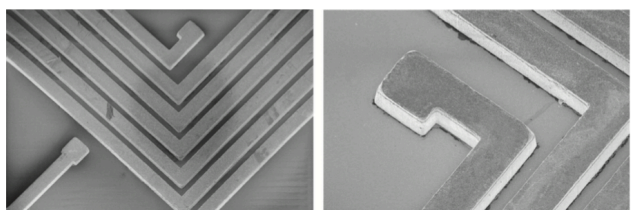
**Fig 12:** SEM image of evaporated micro-coils, reprinted from (*58*).

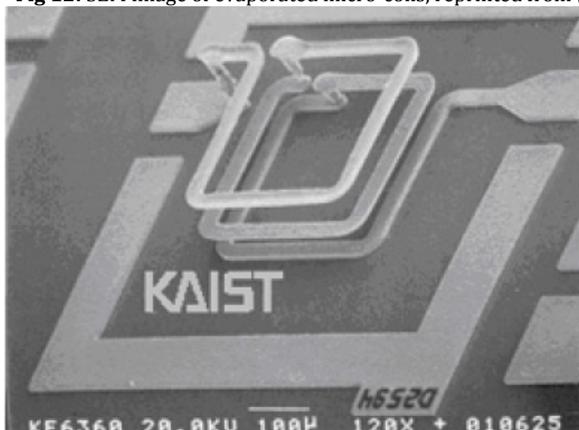
**Fig. 13:** μcoil monolithically obtained in CMOS technology by metal lines, SEM image from KAIST

Plasma cleaning and wet etching is used to remove both traces of organic resists and sticking layer in chromium and gold in order to obtain finally the single coil traces. The passivation (typically in SiO₂) is finally deposited to assure protection of the coil spires and the final structure is shown in figure 13. Wafers may be then cut to separate the inductors and aluminium wire bonding used to connect the inner edge of every inductor to the contact pad laying outside the spiral. Epoxide resin is applied over the bonding area to prevent short-circuits.

A second strategy is instead considered by realizing the microcoil directly integrated with the CMOS circuit by exploiting the several allowed metal layers. This CMOS fabrication is considered in two options: the first option is to build the μcoil monolithically embedded in the CMOS architecture by surrounding the metal lines and vias required to contact the sensing electrodes on the top of the die, as well as in the structure shown in Figure 13.

A further option is instead to use all the metal layers and all the available VLSI area to realize only the coil. Then, in a post-processing step, the chip of the coil is flip and bonded to the other CMOS chip where the sensing system would be independently realized. Both the options are considered and the best one in terms of better system performance and fabrication costs may be chosen for the final demonstrators.

The advantages of the first approach is the degree of freedom in designing and realizing the μcoil in a separate silicon substrate without CMOS constrain (metals to be used, fabrication sizes, etc.), while the advantages of the second approach is the full compatibility with the CMOS process and the simultaneous fabrication of the coil with the CMOS circuit, in the option for the monolithic integration.

Of course, in case that these proposed strategies for powering our CMOS dust by mean of an inductive link will reveal a too-low received energy with respect the power demand by the integrated electronics for sensing and telemetry, then alternative approaches for energy harvesting might be considered and suggested as guidelines for creating sub-10μm CMOS systems for drinkable electronics. For example, alternative approaches such as galvanic processes involving intestine acids (*59*) or energy-recovery by the glucose content in blood streaming (*60*) may be considered.

## System Design: the nanosensors
*Formation of Au electrodes*
Recently, we have already successfully integrated nano-Pt structures directly on top of a CMOS die as well (see Figure 14) and succeeded in using the chip for bio-detection (*22*). By using photolithography-based techniques, we propose to create gold (Au) electrodes above the Aluminium electrodes on the CMOS chip surface. Gold is suitable because it an inert material that is biocompatible, is relatively inert, has a relatively wide electrochemical potential window, and is very commonly used in electrochemical systems since it allows immobilization of biomaterials. Electrodes are constructed by first depositing a 5-nm titanium (Ti) adhesion layer under a 100-nm gold (Au) layer and patterning the layers by using photolithography (*21*).

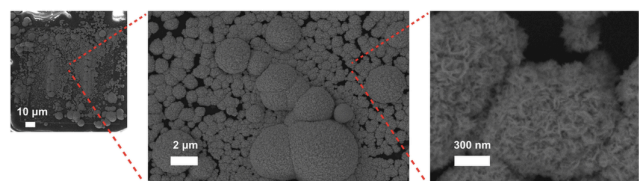
**Fig 14:** SEM images at different magnification of nano-Pt directly growth on the top metal layer of a CMOS die, reprinted from (*22*).

*Formation from tetravalent platinum*
Pt nanopetals with average size in the Nanoscale (of around 68 nm) are produced at the exposed pad-electrodes. The IC electronics is also designed to be capable the right potential for the formation of nanosensors by electrodeposition once the CMOS cube is exposed to a solution containing tetravalent platinum salt. In particular, the chip function will apply −1V for 200s, meanwhile the IC' electrodes are exposed to a solution of 25mM $H_2PtCl_6$ and 50mM $H_2SO_4$ (*38*).

To study alternatives on forming the nanosensors, the influence of the deposition solution is optimized by also considering different concentrations for both the Pt salt (the $H_2PtCl_6$) and the $H_2SO_4$ as well. Ranges of concentrations, e.g. from 3mM to 500mM are usually considered for the case of $H_2PtCl_6$, while ranges from 25mM to 50mM are usually considered for the $H_2SO_4$. The differently formed electroactive areas on top of the IC' electrodes may be then investigated by using the SEM (Scanning Electron Microscopy) in order to verify the quality of the obtained nanosensors and to investigate the different outputs obtained by the different concentrations of species in the forming solutions.

*Formation from divalent platinum*
The self-formation of nanostructured sensor on top of our IC dies are also optimized by using a solution containing a divalent platinum salt: the $K_2PtCl_4$ (again, in solution with $H_2SO_4$). It has been demonstrated that the morphology we can obtain with this different salt may result from nanospheres if applying the voltage of −0.2V. Therefore, we also consider nanostructures by divalent platinum salt by using the same forming IC function that apply a potential of -1V but proposing in this series of experiments the divalent platinum salt, instead of the tetravalent platinum salt used before. Here, we can also suggest to change solution concentrations in ranges from 25mM to 37.5mM for the $K_2PtCl_4$ (*38*) too, and again from 25mM to 50mM for the $H_2SO_4$. Similarly, the differently formed electroactive areas on top of the IC' electrodes may be then investigated by using again the SEM in order to verify the quality of the obtained nanosensors and to investigate the different outputs obtained by the different concentrations of species in the forming solutions. Nanostructures with features from 47 nm to 70 nm are expected by using this methodology based on divalent platinum salt.

*Subsequent Formations*
To further improve the electroactive area of the electrodes, we could also carried out successive depositions, e.g., running the same self-forming functions several times in series. For example, after the deposition of nanospheres (e.g., −1 V, 200 s, 25 mM $K_2PtCl_4$ + 50 mM $H_2SO_4$) we synthesize Pt nanopetals onto the overgrown electrode (e.g., −1 V, 90 s, 25 mM $H_2PtCl_6$ + 50 mM $H_2SO_4$). This output is obtained by using the same forming function of the IC by exposing the CMOS die to two different Pt-salt solutions. Therefore, we then obtain Pt nanopetals grown on nanospheres. The quality of the obtained two differently shaped nanostructures may be, again, investigated by SEM microscopy in order to check and verify the quality of the obtained sensing nano-structured surfaces.

## System Design: the biosensors
The formation of the biosensors on top of the IC dies is obtained by electrochemical deposition (or condensation) of proper polymer that could traps the proteins on its matrix. In fact, the working electrodes of the CMOS dye are individually functionalized with high precision with a single-step electrodeposition by using solutions containing chitosan (*61*). As described in the previous sections of the state of the art, different proteins are used to specifically target different cancer markers, and/or human metabolites, and/or chemotherapeutic compounds. The polarization of the working electrodes at a certain potential for a certain time lap will creates a localized region of high pH that exceed chitosan' solubility limit, allowing the materials polymerization and, therefore, the entrapment with high spatial selectivity of any other compound present in the solution, including the proteins present in the same solution. The variation in the electrodeposition time will allow the control the amount of deposited proteins. Therefore, we expose the top working electrodes of the die to a solution containing proper proteins and the chitosan, and the right function in the IC is capable to trigger the formation of the biosensors on top. To deposit the proteins by chitosan trapping, the forming functions of the CMOS IC will provide a proper potential at the electrodes where the sensing film needs to be formed. The IC is exposed for 5 min to a deposition solution. To optimize the biosensors performance, we consider deposition solution in the ranges of concentration of about 0.5% of chitosan solution and ranging from 10 to 50 μM for the protein. After the electrochemical deposition of the probe proteins, the functionalised CMOS dyes are then rinsed with PBS buffer at pH 7 (to preserve the proteins' function) and stored in the buffer at 4 C before tests. As an alternative, we also propose to consider a different forming procedure with the aim to see if we can find a better method. In this second case, the electrodeposition of proteins is carried out in a pH 5.6 electroplating bath containing also palladium to assure the electroplating and to enhance the final biosensor properties (*62*). The composition of the palladium–protein-electroplating bath consists $PdCl_2$ and the probe protein. By following the originally proposed procedure (*62*), ranges of concentration from 0.5 to 3 mM are considered for the $PdCl_2$ and from 300 to 1500 Units for the protein, in order to optimize the performance of the formed biosensors. In this case, the forming function of the IC provides a constant potential of −0.9 V for 10 min.

## System Design: Biocompatible packaging
Packaging of the system should provide a special coating that is biocompatible whilst allowing a series of special functions: (i) allowing passing through the gut-wall barrier; (ii) targeting a specific tumor mass; (iii) allowing human molecules (tumor biomarkers, metabolites, and anti-cancer chemotherapy compounds) to penetrate the packaging for reaching the sensors. In past, we have already realized and successfully tested several biocompatible packages for fully implantable devices. In particular, we have verified that the use of an epoxy resin is sufficient to the aim. In our experiments with mice, we didn't registered any further inflammation, nor in terms of increased neutrophils neither in terms of increased ATP (adenosine-triphosphate), with respect to commercially available RFID-chip we normally implanted in our pet dogs (*10*). On the other hand, the use of epoxy resin provides a semi-permeable membrane that allows body metabolites to reach the sensing electrode of the diagnostic device. Therefore, the CMOS sensing cube is covered with layers of a membrane made by epoxy-enhanced polyurethane. An already successfully tested procedure is followed (*10*): a homogeneous solution is obtained by mixing 125 mg of an epoxy adhesive (EP42HT-2Med system), purchased by Master Bond (Hackensack, USA) as a certified biocompatible two-components adhesive, 112.5 mg of polyurethane (Sigma Aldrich), 12.5 mg of the surfactant agent polyethylene glycol ether (Brij 30, Sigma Aldrich), for 10 ml of tetrahydrofuran (THF, Sigma Aldrich) is used as solvent. In case, subsequent depositions of the membrane are applied at 1 h intervals and then the CMOS dyes are stored overnight at room temperature. A fast curing at high temperature (2 h at 80 C) is needed to ensure the biocompatibility of the resin. After this process, the CMOS dyes are again kept overnight at room temperature, and then stored in PBS one day for membrane swelling. A successive step of functionalization on the surface of the epoxy resin is then performed in order to transfer proper ligands for targeting the tumour mass. Target molecules that are specific for tumor cells, for example prostate specific membrane antigens or Transferrin receptors (*63*), are considered to identify the related ligands, e.g., small molecules or antibody fragments. The epoxy resin is then functionalized with these ligands by using conventional linkers to bind the ligands to the resin. It is well known that the amine (NH2) groups react with the epoxide groups of the resin during polymerisation (*64*), while the structure of proteins, e.g., an antibody, contains a lot of amino acid residues with exposed amine groups useful to cross-link with functional surfaces (*65*). A final step of functionalization is done to transfer proper lipids into the packaging membrane. This is required by the mechanism of internalization by M and dendritic cells (*12*). Lipid raft domains (*66*), or coatings with lauroyl chains and propranolol molecules (*67*) are both considered to provide an augmented endocytosis. Alternative organic or biological molecules suitable for integration in the epoxy membrane and carrying an internalization function may be identified in literature and, in case, integrated in the packaging as well. Even though it has been never tried (up to our knowledge), a similar biocompatible semi-permeable membrane may be further functionalized with molecules that manifest affinity to specific receptors at the surface of human cells related to a disease in order to target a specific body region. For example, in case of a tumor mass, it is possible to functionalise such a membrane with affinity ligands that target tumor cells as well as it is usually proposed for active targeting in anticancer therapies (*11*). The day the technological progress will succeed in developing CMOS diagnostics tool with size of very few micros, a further function might be transferred to the biocompatible membrane: the capability to trigger the internalization, a process that allow to pass the gut walls thank to Microfold (M) cells and dendritic cells (*12*). The mechanism of internalization is a phenomenon of taking molecules (e.g., an antigen) from the lumen to the intestine (and vice-versa) by means of the endocytosis and exocytosis. Endocytosis and exocytosis are two reversed processes used by cells to engulf molecules and then provide an active transport. In dendritic and M cells, this mechanism also allow the transport of quite large bio-structures, like liposomes that can reach size up to 2.5 μm (*13*). The process of internalisation is also modulated by lauroyl chains and propranolol molecules (*67*), as well as by lipid raft domains that have shown an augmented endocytosis (*66*). Therefore, lipids are suitable organic molecules to be used for transferring further function on biocompatible membranes to support active transports in biological cells.

## Potential Applications and Impact of Technology

In this paper we introduce the concept of Body Dust, a new kind of CMOS diagnostics in the form of dust particles integrating bionanosensors for continuous monitoring of several human metabolites. Such a new technology we are proposing here will have a big social and economic impact on three main socio-economic groups: IC industry (the industry of production of integrated electronics), the biomedical circuits and systems community, and the pharma industry as well.

*Microelectronics industry*
Demand for new bioelectronic systems available in the market will definitely increase in the following years mainly due to an expanding number of end-user applications based on personal electronics (e.g., smartphones, tablets, connected watches, etc). This will provide further stimulus to the growth of the microelectronics market worldwide.
More important, microelectronics post processing becomes a key factor for many of these new applications and, in some cases, it can be cumbersome to preserve the functions of the underneath architectures. Therefore, the global microelectronics industry will benefit of the completely new approach we are proposing in this paper for the CMOS design as well as for the post processing steps. Nowadays, the post processing is typically manufactured for hybrid ICs by depositing and patterning metals after the chip fabrication. Typical example is the post processing required for integrating MEMS sensors (e.g., accelerometers) and the readout electronics. Our new approach, will offer to the global IC industry a completely innovative way to have formed and self-programmed post processes deeply integrated in the CMOS architectures with possibility of easy formation of nanostructured and/or bio-functionalized electrodes on top of the hybrid ICs for many different applications.

*Biomedical Circuits and Systems*
In diagnostics, new concepts such as home diagnostics and remote monitoring are emerging. These new concepts require the commercialization of a completely new class of autonomous devices. This is reflected in the worldwide effort to develop healthcare equipment that support portability, reduced size, low-power consumption, low-costs in fabrication, easy-to-use or fully-autonomous functions. However, this is possible only with an intimate integration with electronic architectures that provide complex functions, such as automatic acquisition, processing, fusion, communication, integration, and control of data. This new emerging market (the worldwide market for consumer devices in healthcare personal diagnostics) is estimated to reach US$47.40 billion by 2020 with an annual growth rates for investments of around 6% in the period 2016-2020. Of course, this market includes sensors for temperature, blood pressure, heart rate, ECG, blood glucose, blood oxygen, posture, altitude, motion, and others. Our new approach will offer the global industry in the sector of the personal diagnostics the possibility to expand to many other detection targets and to obtain integrated and highly-sensitive (thank to nanostructures) diagnostics biochip for different applications.

*Pharmaceutical industry*
A relevant part of the whole world economy is driven by the *pharmaceutical industry*. Worldwide, the pharmaceutical industry reached around US$1 trillion. North America is the largest portion of such market with more than 40%, while new markets, e.g. the Chinese one, show incredibly high growth rates over the last five years. Usually, the pharmaceutical companies invest nearly around 20% of their revenues in research, development, and test of new compounds. Thus, the development of completely new, low-cost, innovative dust devices that could be spread in bioreactors for continuous monitoring in full autonomy mode the production will offer great benefits to worldwide pharma industry in saving money and shortening the time-to-market for their products. This new technology will furnish a further competitive market advantage, business development and raise of credibility to pharma companies and providing them a first-in-the-world technology to monitor their bioreactors with a diagnostic dust just spread in the cell media as well as any other feeding component of the cell culture.

## Conclusions

In this paper, we have introduced the concept of a new CMOS bio-sensing technology, which is so small that it could be drinkable for in-body diagnostics. The realisation has been shown through the design of a CMOS diagnostic chip to form Body Dust, integrating bionanosensors for continuous monitoring of several human metabolites. This relies on all the discussed methodologies to address the several design challenges, which allow such aggressive miniaturisation using state of the art CMOS fabrication. We have also discussed the huge social and economic potential impact of such a new technology.
Three main socio-economic markets, dealing with integrated electronics, biomedical circuits and systems, and pharma, will get significant improvement by this approach. The theoretical basis for such an approach has been described in this paper, while a true demonstration by means of a prototype is still in progress. Furthermore, some aspects of the proposed approach still require more deep analysis, e.g. the power transmission that allows us to realise receiving coils with the size of 10 μm in lateral size. However, we have proposed here the basis of such new approach in diagnostics as well as we have identified the technological challenges which we need to address in order to make it real.

**ACKNOWLEDGMENTS.** The authors would like to thank for the many useful discussions: Ralph Etienne-Cummings, from John Hopkins University, Georges Gielen, from KU Leuven, Danilo Demarchi, Politecnico di Torino, Ravinder Dahiya, University of Glasgow, James R LaDine from Thermofisher, and all the participants to the Special Session the authors have organized at the international IEEE conference BioCAS 2017 about this topic.

---